\documentclass[aps,twocolumn,footinbib,showpacs]{revtex4}
\usepackage{graphicx,amsmath}

\newcommand{\figwidth}{0.8\columnwidth}
\newcommand{\eq}[1]{Eq.~(\ref{#1})}
\newcommand{\fig}[1]{Fig.~\ref{#1}}

\newcommand{\olcite}[1]{Ref.~\onlinecite{#1}}

\newcommand{\bhu}{ {\bf \hat{u}} }
\newcommand{\br}{ {\bf r} }
\newcommand{\bhn}{ {\bf \hat{n}} }

\newcommand{\kbt}{k_{\rm B}T}

\begin{document}

\title{Aggregation of self-propelled colloidal rods near confining walls}
\author{H. H. Wensink}
\altaffiliation{Present address: Department of Chemical Engineering, Imperial College London, South Kensington Campus, London SW7 2AZ, United Kingdom}
\author{H. L\"{o}wen}
\email{hlowen@thphy.uni-duesseldorf.de}
\affiliation{Institut f\"ur Theoretische Physik II: Weiche Materie,
Heinrich-Heine-Universit\"at D\"{u}sseldorf, 
Universit{\"a}tsstra{\ss}e 1, D-40225 D\"{u}sseldorf,
Germany}

\date{\today}

\begin{abstract} Non-equilibrium collective behavior of self-propelled colloidal rods 
in a confining channel is studied using Brownian dynamics simulations
and dynamical density functional theory. We observe an aggregation process
in which rods self-organize into transiently jammed clusters at the channel walls. 
In the early stage of the process, fast-growing hedgehog-like clusters 
are formed which are largely immobile. At later stages, most of these clusters 
dissolve and mobilize into nematized aggregates sliding past the walls.
\end{abstract}
\pacs{82.70.Dd, 05.40.-a, 61.20.Lc, 61.30.-v}

\maketitle

Swimming microorganisms,  insects, birds and fish frequently 
move collectively in large groups
with spontaneous liquid crystalline order. Considerable recent research activity
has been devoted to understand the origin of flocks and swarms 
in terms of simple models
of self-propelled particles \cite{Vicsek,Tonertu,Levine,Ramaswamy,Orsogna,Bar,Chate,Saintillan}. 
In these models, ``active'' rods are driven
by their own motor along the rod orientation axis and dissipate energy in the suspending medium. In fact, it has been shown that a local interaction between 
neighboring particles which
favors mutual alignment induces a non-equilibrium phase transition from a state of 
zero transport to another one involving a big swarm with cooperative 
transport 
\cite{Vicsek}. This nematically ordered state has been analyzed with 
hydrodynamic approaches \cite{Ramaswamy}
and instabilities at large wavelengths and giant density 
fluctuations were predicted
\cite{Ramaswamy,Chate,Saintillan}. In a model with explicit rod 
interactions, self-organized vortex clusters were found 
in which particles rotate around a common center \cite{Levine,Orsogna}.

Recently  self-propelled rods have been realized in fluidized
monolayers of macroscopic rods \cite{Narayan} and by means of vibrating
asymmetric granular  rods  \cite{Kudrolli}. Giant density fluctuation 
were confirmed \cite{Narayan}. In these realizations,
the rod-rod interaction is the crucial input in contrast to collective
motions of bacteria or  low-Reynolds-number 
micro-swimmers  \cite{Sokolov} which are dominated by
hydrodynamic interactions  \cite{marenduzzoyeomans,Pagonabarraga}.
While most investigations have been performed in the bulk, 
mainly in two spatial dimensions (2D), there are few studies of 
self-propelled rods
in confinement. Micro-swimmers moving in channels
have been investigated in \olcite{Hernandez}. Self-propelled granular rods 
in circular-shaped cells were studied experimentally  in \olcite{Narayan} 
and clustering of particles near the cell boundaries was found  \cite{Kudrolli}.

In this paper we study the effect of channel  confinement on the 
collective
behavior of self-propelled interacting colloidal rods
 by means of Brownian dynamics computer simulation and  dynamical density functional theory.
Our motivation to do so is threefold:
First, since every realization in nature is finite, there is a general need
to clarify the collective behavior of active rods close to system boundaries,
and, in particular, to feature the transport properties of colloidal particles.
Second,  transport through narrow channels is omnipresent
in many realizations, e.g. the propagation of microorganisms through veins and pores.
Third, a predictive theory which starts on the level of the interactions 
and incorporates microscopic correlations is needed. 

A 2D slit will cause the rods to align
along the channel walls such that a  nematized steady state with a
local nematic director pointing along the wall direction is expected for finite densities. 
However, for active particles, we observe a novel  clustering
phenomenon close to the walls occurring on top of a nematization. We find an aggregation
of rods at the channel walls into transiently jammed clusters  which
exhibit a {\it hedgehog-like} structure. 
Most of these immobile ``hedgehog'' clusters show up 
in the early stages of the aggregation process. They
can efficiently block
particle transport through the channel. 
The clusters do not only form in 2D channels 
but also appear in thin capillary slits in 3D. For soft particle interactions
considerable homeotropic order near the walls is found which is purely generated by non-equilibrium.
The aggregation
near  walls and the stability of hedgehog clusters is also borne 
out by a dynamical density functional theory 
which we construct from the Smoluchowski 
equation \cite{Dhont,Nagelephysrep,Baskaran}.

 The clustering process can 
be detected in experiments using catalytically driven nanorods 
\cite{Bala, Dhar} and colloidal particles \cite{Erbe}
in micro-channels which exhibit the Brownian dynamics of our model \footnote{Due to the solvent, the dynamics is different from granulates where inelastic
 collisions are important, see I. S. Aranson and L. S. Tsimring, Phys. Rev. E {\bf 71},  050901 (2005)}.
A control over the hedgehog cluster formation has relevant implications for 
 transport of active particles through confinements.

In our model, we describe a 2D system of $N$ charged Brownian rods of 
length $L_{0}$ confined in a channel  with a fixed width $\Delta x = 5L_{0}$. 
Rod-rod interactions are implemented via a segment 
model where each rod is partitioned into 13 equidistant segments \footnote{Qualitatively similar results were obtained for 20 segments per rod.}. Each two 
segments from different rods interact with each other via a 
Yukawa potential \cite{kirch-bd}. The pair potential between two segmented rods at  given 
center-of-mass positions $\{ \br , \br ^{\prime} \}$ and orientational unit vectors
$\{ \bhu , \bhu ^{\prime} \}$  is then given by
\begin{equation}
U_{\text{rod}}(\br - \br ^{\prime} ; \bhu ,\bhu ^{\prime} ) = U_{0} \sum 
_i \sum _j \frac{\exp [ -\kappa r_{ij} ] }{  \kappa r_{ij}},
\end{equation}
with $\kappa$ the Debye screening constant and $r_{ij} = | \br - \br ^{\prime} + (\ell_{i} \bhu - \ell_{j} \bhu ^{\prime}) |$ 
the distance between segment $i$ and $j$ with $ -L_{0}/2 < \ell_{i/j} < L_{0}/2$. In all cases, $U_{0} = 20$ 
$\kbt$ with $k_{\rm B}$ Boltzmann's constant and $T$ temperature. The 
same Yukawa segment potential is used to model the rod-wall interaction:
\begin{equation}
 U_{\text{wall}}( x , \bhu  ) = U_{0} \sum _i  \frac{ \exp [-\kappa x_{i} ] }{  \kappa x_{i} },
\end{equation}
with $x_{i} = |x + \ell_{i} \bhu |$ and $x$ the rod center-of-mass  distance to the channel wall. 
This potential prevents any particle-wall overlap and enforces the rods to orient along the channel direction.
\begin{figure} 
\begin{center}
  \includegraphics[clip=,width=  \figwidth ]{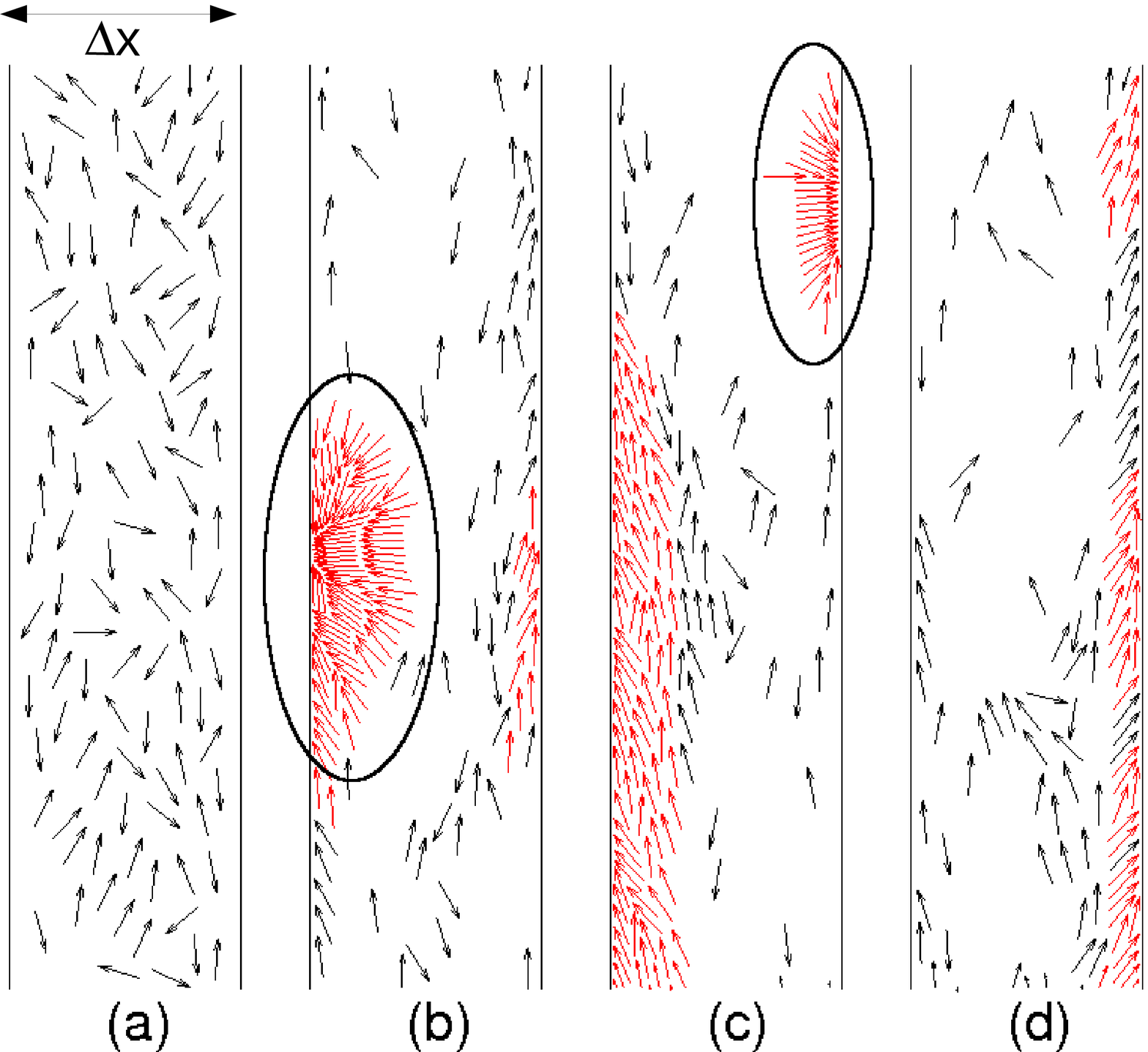} 
\includegraphics[clip=,width=  \figwidth ]{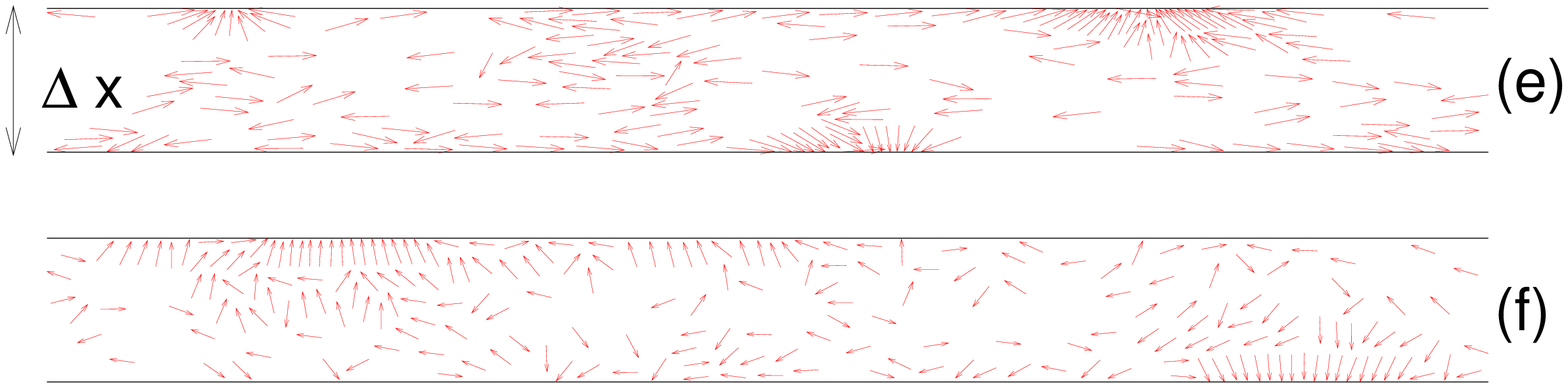} 
\caption{\label{snap} (Color online) (a-d): Simulation snapshots (cutout of the full simulation box)
showing the aggregation process  in a 2D system of 
 self-driven rods with $F_{\parallel}=5\kbt/L_{0}$, $\kappa L_{0} =10$. (a) 
$t=0$ (isotropic), (b) $t=0.10\tau_{B}$ (hedgehog), (c) 
$t=0.18\tau_{B}$ (nematized hedgehog) and (d) $t=0.73 \tau_{B}$ (wall aggregates). Clustered rods are shown in red, unclustered ones in black, typical hedgehog clusters are encircled.  Arrows indicate the direction of $F_{\parallel}$. (e-f): Wall clusters appearing in a 3D {\em  slit capillary} with width $\Delta x=5L_{0}$ and height $\Delta z =0.3 L_{0}$ at the same driving force. Shown are the projections onto the $xy$-plane. The parameters are $\kappa L_{0} = 10 $, $NL_{0}^3/V=5$ (with $V$ the system volume), $ t = 0.09 \tau_{B} $ (e) and $\kappa L_{0} = 5$, $NL_{0}^3/V=3$, $ t = 0.15 \tau_{B}$ (f).} 
\end{center} 
\end{figure}

\begin{figure}
\begin{center}
\includegraphics[clip=,width=\figwidth ]{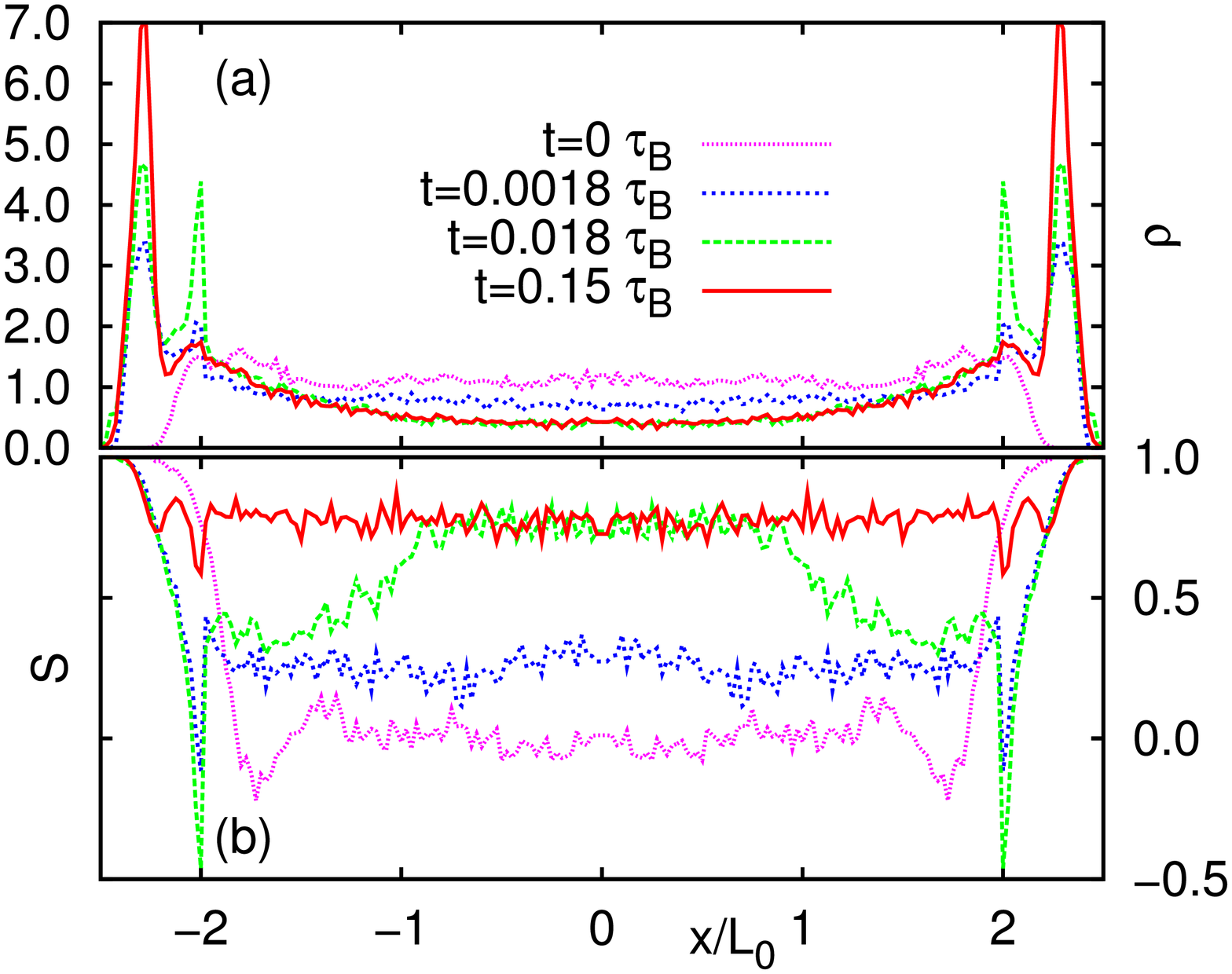}
\includegraphics[clip=,width=\figwidth ]{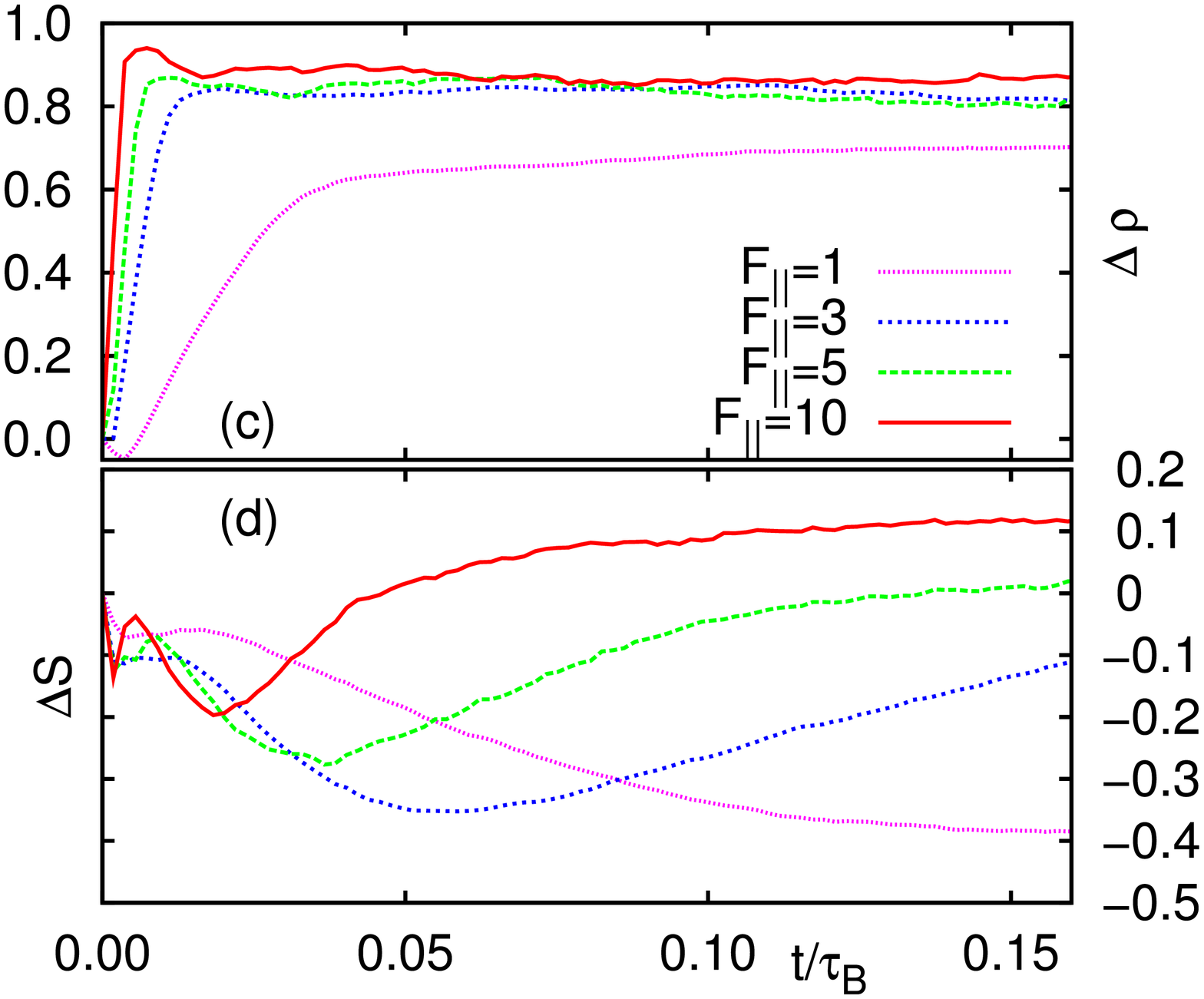}
\caption{\label{sims} (Color online) Time-dependent profiles for the number density $\rho(x)$ (a) and 
nematic order parameter $S(x)$ (b) for $F_{\parallel}=10\kbt/L_{0}$, 
$\kappa L_{0} =10$ obtained from simulation. (c) Adsorption $ \Delta \rho = \int _{0} ^{L_{0}} [ \rho 
(x,t)- \rho (x,0) ] dx $ and (d) excess orientation $ \Delta S = \int _{0} 
^{L_{0}}\rho(x) [ S (x,t)- S (x,0) ] dx / \int _{0} ^{L_{0}} dx \rho(x) $  showing the 
evolution of the rod structure with respect to the equilibrium initial state near the channel walls.}
\end{center}
\end{figure}

To model an ensemble of self-propelled rods a constant force
$F_{\parallel}$ is applied along the main axis of each rod.
The system is kept at 
constant temperature $T$ via the solvent and the dynamics is 
simulated by a finite-difference integration of the overdamped Langevin 
equations using the short-time translational and rotational diffusion 
constants of a rod taking the limit of infinite aspect ratio \cite{kirch-bd}. At any finite number density, the packing fraction of the system is 
then negligibly small and hydrodynamic interactions between the rods are of minor importance. 
The overall number density  $ \rho_{0} = NL_{0}^2/A$ (with $A$ the system area) is fixed at $\rho_{0}=1$ which corresponds to a 2D isotropic equilibrium bulk structure. 
Time $t$ is measured in units of the Brownian relaxation 
time $\tau_{B}=L_{0}^2/D^{\parallel}_T$ of a single rod in terms of the translational diffusion coefficient parallel to the 
rod $D^{\parallel}_{T}$. All simulations were carried out with a 
  time step $\delta t = 10^{-7}\tau_{B}$, with a periodically repeated finite
system size of $125L_0$ along the channel comprising $N=625$ rods. 
Rod clusters were identified using a distance criterion such that a rod pair is assigned to a cluster 
if their distance of closest approach is less than $ \kappa^{-1}$.

In all processes, the initial rod configuration is the isotropic 
equilibrium state for $F_{\parallel}=0$. At $t=0$, $F_{\parallel}$ is switched to a 
finite value with the forces pointing randomly along the main rod axis, as depicted in \fig{snap}a,
and  rods start accumulating at the walls forming 
tight clusters with a large fraction of rods pointing along the wall normal (\fig{snap}b).
 Owing to its 
symmetric, semicircular shape these wall clusters will be referred to as {\em hedgehogs}. Since the net 
self-propulsion force is very small the hedgehogs are practically immobile which enables them to gather many other rods
and grow rapidly. When the hedgehog has reached a certain critical size it becomes 
structurally unstable and slowly dissolves into a  nematized 
cluster with nonzero cooperative motion (\fig{snap}c). Simultaneously, new hedgehogs may form at a vacant part of the wall. 
In the steady state (\fig{snap}d), the nematized hedgehogs have spread 
out along the walls and the rods move cooperatively in mono-layer aggregates 
displaying a characteristic tilt angle with respect to the wall and 
only few hedgehog clustering events are seen.
Steady state density fluctuations are measured from 
$\sigma ^2  = \langle  M ^ 2 \rangle - \langle M \rangle ^2 $ with  $M$ the number of rods in the left (L) or upper (U) half part of the simulation box. 
The ratios, $\sigma _{L} /\sigma _{L,eq} = 11 $, $\sigma_{T}/\sigma _{U,eq}= 20$  (for $F_{\parallel}=10\kbt/L_{0}$) confirm huge density fluctuations for confined driven systems. 

 A  simulation study in a 3D slit geometry (see \fig{snap}e) shows that the formation of hedgehogs is
 robust  against the rods moving out of the 2D plane.
In  \fig{snap}f we observe that a reduction of the effective aspect ratio $\kappa L_0$ leads to a transformation
of the hedgehogs  into  {\it linear homeotropic}
 wall clusters consisting of arrays of parallel oriented rods. 
These type of clusters appear to be persistent in time, also in a 3D slit-geometry. Note that the homeotropic alignment is a purely  non-equilibrium effect which is not imparted by the external wall potential.

\begin{figure} 
\begin{center} 
\includegraphics[clip=,width= \columnwidth ]{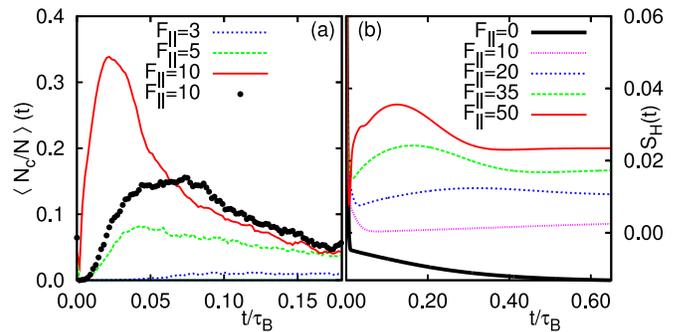} 
\caption{\label{clus} (Color online)  (a) Average number fraction of 
clustered rods ($N_{c}/N$) versus time. The solid curves correspond to an 
initial state of freely rotating rods, the points to an aligned state 
where $ \bhu _{i} \perp \bhn $ for each particle $i$. (b) Time evolution of the hedgehog strength ${\cal S}_{H}(t) $ of a hedgehog nucleus (see \eq{nucleus}) from dynamical density functional theory. }
\end{center} 
\end{figure}

The behavior depicted in  \fig{snap}a-d 
also globally shows up in the profiles for the density $\rho(x)$ 
and nematic order parameter $S(x)$ across the slit, shown in \fig{sims}.
The latter is defined as  
\begin{equation}
S(x) = \frac{1}{N} \sum_{i} \langle 1 - 2 ( \bhu _{i} \cdot \bhn )^2  
\rangle _x, 
\end{equation}
with $\langle \cdot \rangle _x $ denoting a 
local canonical average at position $x$. $S$ is close to unity if rods are preferentially parallel to the wall, 
whereas a negative value indicates a favorable perpendicular alignment. 
 The density profiles are obtained by a time-dependent canonical average 
of the microscopic density using about 500 independent 
processes. The sharp increase of the wall peak points to a strong 
aggregation of rods at the wall. The double peaked shape of $\rho(x)$ at 
intermediate times can be attributed to the presence of both hedgehogs 
(associated with a sharp minimum in $S(x)$) and nematized wall clusters. 
Details of the aggregation process are depicted in \fig{sims}c-d.  There, 
the rod structure in the direct vicinity of the walls is shown in terms of 
the integrated difference of the profiles with respect to the equilibrium 
state at $t=0$.  The excess orientation (\fig{sims}d) varies 
non-monotonically with time and the minimum can roughly be identified with 
the point where the transient hedgehog clusters break up and transform into 
nematized ones (see \fig{snap}). From this we can then infer that: (1) the 
average lifetime of the hedgehogs becomes smaller upon increasing 
$F_{\parallel}$ and (2) the ones formed at smaller driving forces have an 
increased propensity to be perpendicular to the wall.
A similar non-monotonic behavior is encountered in the average cluster size as a function of time shown in 
\fig{clus}. For the largest forces, the maximum appearing in the cluster 
fraction roughly corresponds to the ones in \fig{sims}d which provides 
additional support for the transient clustering scenario. 

We stress that at later times, say $t>0.1\tau_B$, hedgehogs do still
appear (see \fig{snap}c) albeit with a smaller probability.
To ensure that the qualitative features of the clustering scenario do not 
dependent upon the initial structure, we also considered clustering 
starting from an equilibrated nematic system where all rods are 
enforced to be parallel to the wall. The result, included in \fig{clus}, 
shows that a similar clustering process, albeit less pronounced, takes place starting from 
a collection of nematic swimmers. 
\begin{figure}
\begin{center}
\includegraphics[clip=,width=\figwidth ]{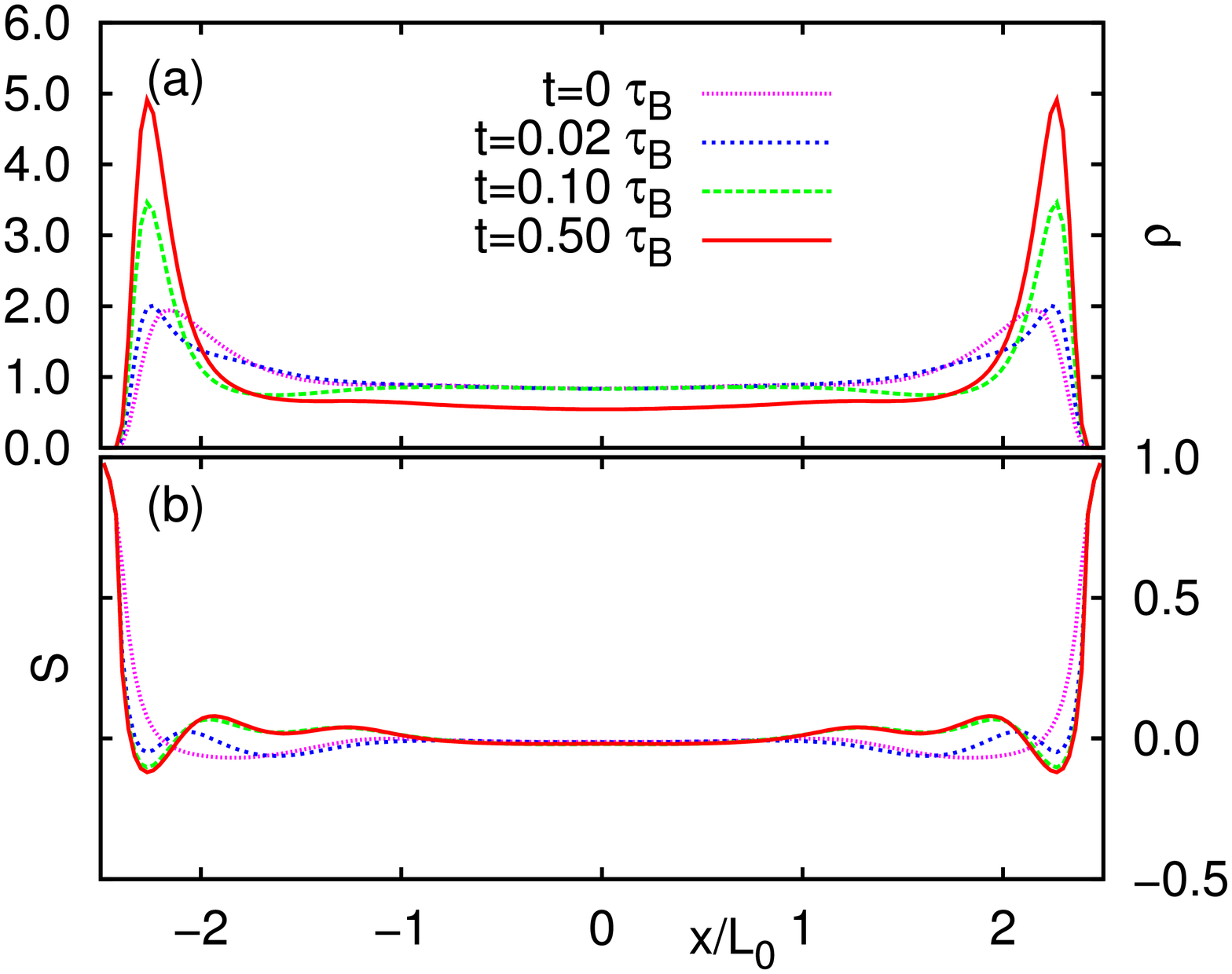}
\includegraphics[clip=,width=\figwidth ]{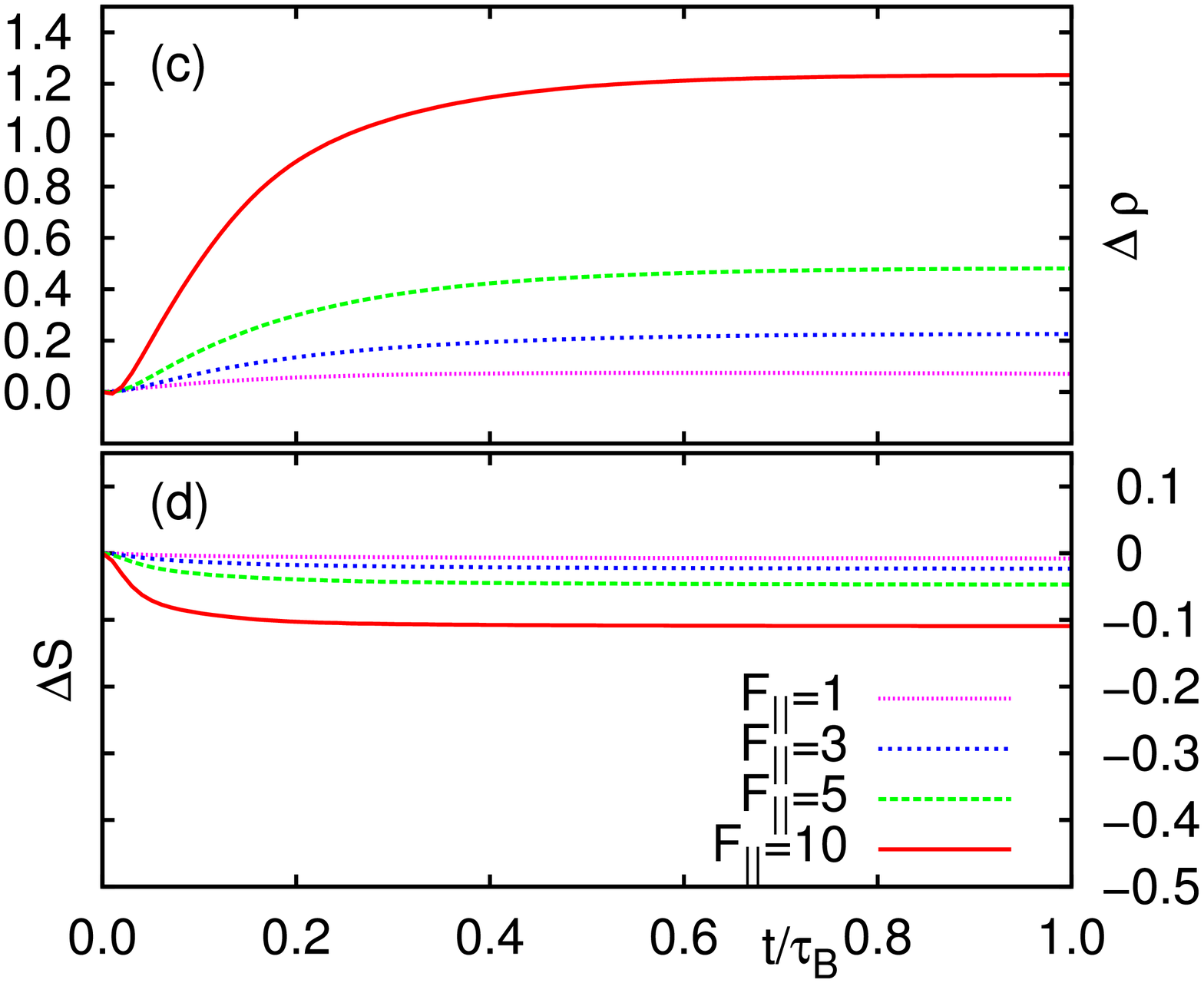}
\caption{\label{theo} (Color online) Same as \fig{sims}. Results from dynamical density functional theory.} 
\end{center} 
\end{figure}

Let us now turn to a microscopic theory for the aggregation process which is formulated
in terms of the one-body density $\rho(\br,\bhu,t)$ for a non-equilibrium ensemble of self-propelled
rods. A general equation of
motion for $\rho $ is obtained by integrating the $N$-particle Smoluchowski
equation which leads to \cite{dhontbriels,rexwensink,Baskaran}:
\begin{eqnarray}
 \partial _t \rho  &=&  \nabla \cdot  {\bf D}_{T} \cdot \left [ \nabla \rho  + \rho \nabla  \left ( \beta U_{\text{int}} +  \beta U_{\text{wall}} \right )  - \beta F_{\parallel} \rho \bhu  \right  ] \nonumber \\
 &+& D_{R}  \left [   \partial_{\varphi}^2 \rho  + \partial _{\varphi} \rho \partial _{\varphi} \left( \beta U_{\text{int}} +  \beta U_{\text{wall}} \right ) \right ], \label{ddft}
\end{eqnarray} 
where  $\beta = (\kbt)^{-1}$, $\nabla = \{ \partial_x, \partial _y \} $ and $\bhu = \{ \sin \varphi, \cos \varphi \}$ with $\varphi$ the rod-wall angle. Furthermore,  ${\bf D}_{T}$ is the diagonal 
translational diffusion tensor and $D_{R}$ the rotational diffusion
coefficient. The effective aspect ratio is fixed at $\kappa L_{0} =10$.
A connection with density functional theory is made by
defining the non-equilibrium collective interaction potential
$\beta U_{\text{int}} = \delta \beta  F_{\text{exc}}[ \rho ] / \delta
\rho $ in terms of the {\em equilibrium} excess free energy functional
$F_{\text{exc}} $ \cite{rexwensink}. Here we use a simple Onsager functional \cite{onsager,poniholy88}
\begin{equation}
\beta U_{\text{int}} ( \br , \bhu ) = - \int d \br ^{\prime} d \bhu ^{\prime} \left (  \exp [ - \beta U_{\text{rod}} ]  - 1 \right ) \rho ( \br ^{\prime}, \bhu ^{\prime}  ),
\end{equation}
which accounts for the rod interactions on the second-virial level.

To describe the wall aggregation process we assume the density  to be variable only along the wall normal $\bhn$ and the one-body densities  for $t > 0$ were numerically obtained from \eq{ddft}.
From \fig{theo} we see that the theory  captures the initial stages of the aggregation process. The
growth of the  wall density peak signifies a  strong adsorption at the
wall while the  concomitant decrease of $S(x)$ points  to the rods tilting
away  from   the  wall.

To ascertain the stability of hedgehogs, we have solved \eq{ddft}  with full 2D spatial resolution focusing on the fate of a small semicircular hedgehog wall nucleus with radius $R_{c}$ at $t=0$ parametrized as follows:
\begin{equation}
 \rho (\br,\varphi , 0) =
  \begin{cases}
   \tilde {\rho }  ^{eq} (x)  \frac{ \exp [  \alpha \cos ( \varphi - \gamma(\br) )] }{ 2 \pi I_{0}(\alpha )} & r<R_c \\
   \rho^{eq} (x,\varphi) & r \geq R_c,
  \end{cases} \label{nucleus}
\end{equation}
in terms of the local director angle $\gamma (\br) =  \arccos ( y/r  ) + \pi $, 
 the equilibrium density distribution $\rho^{eq}$ [with 
$ \tilde {\rho} ^{eq}(x) = \int d \varphi \rho ^{eq} ( x, \varphi) $] and 
local nematic order parameter $\alpha = 100$ ($I_{0}(\alpha )$ is a modified Bessel function). 
The hedgehog {\em size} can be measured in terms of the following weighted density
\begin{equation} 
{\cal R}(t) = \int d \br d \varphi  \exp[-(r/R_c)^2] \rho(\br, 
\varphi, t),
\end{equation}
with $r$ the distance from the center of the hedgehog nucleus at the wall.
Similarly, the hedgehog {\em order} follows from 
\begin{equation}
 {\cal H}(t) = \int d \br d \varphi \cos (\varphi - \gamma( \br) ) \rho (\br,\varphi, t). 
\end{equation}
We may now define the hedgehog {\em strength} ${\cal S}_{H}(t) = {\cal R H}(t)$ as a suitable parameter to assess the stability of the hedgehog nucleus over time. The result in \fig{clus}b shows that a small hedgehog of size $R_{c}=L_{0}$ is indeed stabilized provided that 
the self-propelling force is sufficiently large. Both the transient nature of the hedgehogs and the location of the cluster maximum as a function of the force are in agreement with the simulation results in \fig{clus}a.

In conclusion, we have shown that self-propelled rods confined in a 
channel exhibit an aggregation at the system boundaries. The aggregation
process is dominated by a transient formation of virtually immobile
clusters which possess a hedgehog structure. The results found in 
non-equilibrium Brownian dynamics computer simulations are confirmed by
a microscopic dynamical density functional theory. Our predictions
are in principle verifiable in experiments on 
 colloidal suspensions \cite{Dreyfusbibette}
or catalytically driven nanorods \cite{Bala, Dhar}.
All these experimental systems are essentially realized in 2D. 
The aggregation phenomenon is important for blocking collective motion
of active particles through narrow pores as encountered in microfluidic
devices.

We thank S. Ramaswamy, S. van Teeffelen and M. Rex  for helpful
discussions. This work is supported by the DFG within SFB-TR6 (project D3).

\bibliography{felix}
\bibliographystyle{apsrev}

\end{document}